\documentclass[aps,prl,twocolumn,showpacs,preprintnumbers,amsmath,amssymb]{revtex4}

\usepackage{bm}

\usepackage{epsfig,amsopn}

\usepackage{graphicx}

\usepackage{amsmath,amssymb}

\usepackage{natbib}

\renewcommand{\section}[1]{{\par\it #1.---}}

\newcommand{\beq}{\begin{equation}}

\newcommand{\eeq}{\end{equation}}

\newcommand{\bes}{\begin{subequations}}

\newcommand{\ees}{\end{subequations}}

\newcommand{\bea}{\begin{eqnarray}}

\newcommand{\eea}{\end{eqnarray}}

\newcommand{\ba}{\begin{array}}

\newcommand{\ea}{\end{array}}

\newcommand{\beqn}{\begin{eqnarray*}}

\newcommand{\eeqn}{\end{eqnarray*}}

\newcommand{\f}[2]{\frac{#1}{#2}}

\newcommand{\g}{\gamma}

\newcommand{\al}{\alpha}

\newcommand{\om}{\omega}

\newcommand{\G}{\Gamma}

\newcommand{\la}{\langle}

\newcommand{\ra}{\rangle}

\newcommand{\mT}{\mathcal{T}}

\def\nn{\nonumber}

\def\etal{\emph{et al}}

\begin{document}

\title{Role of pinning potentials in heat transport through disordered
harmonic
chain}

\author{Dibyendu Roy and Abhishek Dhar}

\affiliation{$^1$Raman Research Institute, Bangalore  560080, India}

\date{\today}

\begin{abstract}
The role of quadratic onsite pinning potentials on determining the size
$(N)$ dependence of the disorder averaged steady state heat current
$\la J \ra$, in a 
isotopically disordered harmonic chain connected to stochastic heat
baths, is investigated. For two models of heat baths, namely white noise
baths and Rubin's model of baths, we find that the $N$ dependence of
$\la J \ra$ is the same and depends on 
the number of pinning centers present in the chain.  In the absence of
pinning, $\la J \ra_{Fr}  \sim 1/N^{1/2}$ while  in presence of
one or two pins $\la J \ra_{Fi}
 \sim 1/N^{3/2}$. For a  finite $(n)$ number of pinning
centers with  $ 2 \le n \ll N$ we provide heuristic arguments and
numerical evidence  to show that 
$\la J \ra_{n} \sim 1/N^{n-1/2}$. 
We discuss the relevance of our results in the context of
recent experiments.
\end{abstract}


\maketitle

Since the seminal paper of Anderson \cite{Anderson58}, the physics of
localization in disordered systems has now been studied for over half a 
century \cite{Mott61,Matsuda70,John83,John87}.
Recently there  has been a renewed interest in this field with a lot
of work on some open questions such as, for example, the effect of
interactions on localization
\cite{basko06,pikovsky07,DharLeb08}, 
and the metal-insulator transition in two dimensions \cite{abrahams01}. 
A number of recent experiments have also reported  detailed studies
on localization in varied systems such as heat conduction in a
isotopically disordered nanotube \cite{chang06}, electrons in a disordered
carbon nanotube \cite{gomez05}, photons in a waveguide \cite{lahini08}
and sound localization in elastic networks \cite{hu08}.  The field is
thus still filled with interesting questions and puzzles.
Here in this paper we point out that even the simple problem of heat conduction in a
one dimensional disordered harmonic lattice has surprises. 

It is well known that all the eigenstates of an electron in a
one-dimensional disordered potential are localized. The electrical
current thus decays exponentially with wire length, making it an insulator. In contrast, in
phononic systems, for example a disordered harmonic chain, long
wavelength modes are extended and can conduct a significant amount of
heat. How good a heat conductor then is the disordered harmonic
chain? The obvious question to ask is the system size $(N)$ dependence of
the disorder averaged steady state heat current which we will denote
by $\la J \ra$. It is expected that this
has the form $\la J \ra \sim 1/N^{1-\alpha}$ so that the conductivity scales
as $\kappa \sim N^\alpha$. The dependence of $\alpha$ on choice of
heat baths and boundary conditions has been somewhat puzzling and has
caused some amount of confusion.   
We note that heat conduction in this system is non-diffusive and
correspondingly $\alpha \neq 0$. 

We briefly review earlier work on this problem \cite{LLP03}. 
In an important work  on the localization of normal
modes in the isotopically disordered harmonic chain (IDHC), Matsuda
and Ishii \cite{Matsuda70} (MI) showed that normal modes with frequencies
$\om \stackrel{<}{\sim} \om_d$ were extended. 
For a harmonic chain of length $N$, given the average mass $m=\langle
m_l \rangle$, the variance 
$\sigma^2 =\langle (m_l- m )^2\rangle$ and interparticle spring constant $k$,
it was shown that 
\bea
\om_d \sim \left( \f{k m}{N \sigma^2} \right)^{1/2}
\label{dem}
\eea
They also evaluated expressions for thermal
conductivity of a finite disordered chain connected to (a) white
noise baths and (b) baths modeled by semi-infinite ordered harmonic
chains (Rubin's model of bath). 
In the following we will also consider these two models of baths and
refer to them as model(a) and model(b). 
For model(a) MI  used fixed boundary conditions (BC) and the limit
of weak coupling to baths, while for case (b) they considered free BC and  
this was treated using the Kubo formalism. They found $\alpha=1/2$ in both
cases, a conclusion which we will show is incorrect. 
The other two important theoretical papers on heat conduction in the disordered
chain are those by Rubin and Greer  \cite{Rubin71} (RG)who considered
model(b) and of Casher and Lebowitz \cite{Casher71} (CL) who
used model(a) for baths.
RG obtained a lower bound $\la J \ra \geq 1/N^{1/2}$ and gave numerical
evidence for an exponent $\alpha = 1/2$  
and this was later proved rigorously by Verheggen \cite{verheggen79}. On the other hand,
for model(a), CL found a rigorous bound $\la J \ra \geq 1/N^{3/2}$ and
simulations by 
Visscher with the same baths supported the corresponding exponent
$\alpha = -1/2$. In 
a more recent work \cite{dhar01}, one of us (AD) gave a unified
treatment of the 
problem of heat conduction in disordered harmonic chains connected to baths
modeled by generalized 
Langevin equations and showed that models(a,b)
were two special cases. An efficient numerical scheme was proposed and 
used to obtain the exponent $\alpha$ and it was established that 
$\alpha=-1/2$ for model(a) (with fixed BC) and $\alpha=1/2$ for
model(b) (with free BC).  
It was also pointed out that in general, $\alpha$ depended on the
spectral properties of the baths.

Here we apply the same formulation as developed in
\cite{dhar01} to understand in detail the role of BCs' 
(and more generally the presence of pinning potentials)  on
heat transport in the IDHC connected to either white noise [model(a)] or Rubin 
baths [model(b)]. We show that {{with the same kind of pinning, 
the exponent $\alpha$ is the same for the two different bath
models}}. The pinning potentials strongly scatter low frequency waves
and hence can be expected to lower the heat current. Surprisingly, we
find that even the exponent $\alpha$ changes with the number of
pinning centers.  
We also provide expressions  for the asymptotic value of  $ \la J \ra$
for various cases.

The Hamiltonian of the IDHC considered here is
\begin{eqnarray}
H&=&\sum_{l=1}^N \f{p_l^2}{2 m_l} \
+\sum_{l=1}^{N-1}\f{1}{2}k(x_{l+1}-x_l)^2
\nn\\
&+& \f{1}{2}k'(x_1^2+x_N^2) ~,\label{ham}
\end{eqnarray}
where $\{x_l,p_l\}$ denote the displacement and momentum of the
particle at lattice 
site $l$. The random masses $\{m_l\}$ are chosen from
a uniform distribution between $(m-\Delta)$ to $(m+\Delta)$. The strength
of onsite potentials at the boundaries is 
$k'$. The particles at two ends are connected to  heat baths at
temperature $T_L$ and $T_R$. The heat reservoirs 
are modelled by generalised Langevin equations
\cite{dhar01,oconnor74,dhar06}. The steady state classical heat current
through the chain is given by:   
\bea
J&=&\f{k_B(T_L-T_R)}{4 \pi}\int_{-\infty}^\infty d \om
\mT_N(\om),\label{ocur1}  \\ 
{\rm where}~~
\mT_N(\om)&=&4 \G^2(\omega)|G_{1N}(\om)|^2,~\hat{G}(\om) =
\hat{Z}^{-1}/k \nn \\ 
{\rm and}~~ \hat{Z}&=&[-\omega^2 \hat{M} +\hat{\Phi} - \hat{\Sigma}(\omega)]/k~,\nn
\eea
where $\hat{M}$ and $\hat{\Phi}$ are respectively the mass and force
matrix for the harmonic chain and $\hat{G}$ is the Green's function of
the chain connected to baths. The self-energy correction in the
Green's function  $\hat{\Sigma}$, coming from the baths, is a $N
\times N$ matrix whose only non-zero elements are
${\Sigma}_{11}={\Sigma}_{NN}=\Sigma(\omega)$ and
$\Gamma(\om)=Im[\Sigma]$.   
For white noise baths $\Sigma(\omega)=-i\g\om$ where $\g$ is the
coupling strength with the baths, while
in case of Rubin's baths $\Sigma(\omega)=k\{1-m \om^2/2k-i\om (m/k)^{1/2}
 {[1-m \om^2/(4k)]}^{1/2}\}$. We have assumed that the RG bath 
has spring constant $k$ and equal masses $m$. We
 note that $\mT_N(\om)$ is 
the transmission coefficient of phonons through the disordered
chain. To extract the asymptotic $N$ dependence of 
$\la J \ra$ we need to determine the  Green's function element
$G_{1N}(\om)$. It is convenient to write the  matrix elements
$Z_{11}=-m_1 \om^2/k +1+k'/k-\Sigma/k=-m_1 \om^2/k +2- \Sigma'$ where
$\Sigma'=\Sigma/k-k'/k+1$ and similarly $Z_{NN}=-m_N \om^2/k +2 
-\Sigma'$. Following the  techniques used in \cite{Casher71,dhar01}   we have 
\bea
|G_{1N}(\om)|^2&=& k^{-2}|\Delta_N(\om)|^{-2}~~{\rm with} \label{green}\\
\Delta_N(\om)&=&D_{1,N}-\Sigma' (D_{2,N}+D_{1,N-1})+{\Sigma'}^2 D_{2,N-1}\nn
\eea
where $\Delta_N(\om)$ is the determinant of $\hat{Z}$ and
the matrix elements $D_{l,m}$ are given by the following product of 
$(2\times2)$ random matrices $\hat{T}_l$:
\bea
&& \hat{D}=\left( \begin{array}{cc}
 D_{1,N} & -D_{1,N-1} \\ D_{2,N} & -D_{2,N-1} \end{array} \right)=\hat{T}_1
\hat{T}_2....\hat{T}_N \label{trans} \\ 
{\rm{where}}~~ &&\hat{T}_l=\left( \begin{array}{cc}
2-m_l \om^2/k & -1 \\ 1 & 0 \end{array}
\right) \nn
\eea
We note that the information about bath properties and boundary
conditions are now contained entirely in $\Sigma'(\om)$ while
$\hat{D}$ contains the system properties. It is known
that $|D_{l,m}|\sim e^{c N \om^2}$ for $|l-m|\sim N$ \cite{Matsuda70}, where $c$ is a
constant, and so we 
need to look only at the low frequency ($\om
\stackrel{<}{\sim}1/N^{1/2}$) form of $\Sigma'$ . We now proceed to
examine various cases.  
For model(a) free BC correspond to  $k'=0$ and so 
$\Sigma'=1-i\g \om/k$ while for model(b) free boundaries correspond
to $k'=k$ and this gives, at low frequencies, $\Sigma'=1-i(m/k)^{1/2}
\om$. Other values of $k'$ correspond to pinned boundary sites with an
onsite potential $k_o x^2/2$ where $k_o=k'$ for model(a) and
$k_o=k'-k$ for model(b). The
main difference, from the unpinned case, is  that now $Re[\Sigma'] \neq
1$. The arguments of \cite{dhar01} then immediately give $\al = 1/2$
for free BC and $\al=-1/2$ for fixed BC for both bath
models. The arguments consisted of two parts: (i) it was observed
numerically that the transmission coefficient at low frequencies for
the ordered and disordered chain were almost the same, (ii) an
asymptotic analysyis was then carried out for the ordered case, for
which $\mathcal{T}$ could be obtained exactly for any bath spectral
properties (an improved version of those arguments is given below).

\begin{figure}[t]
\begin{center}
\includegraphics[width=8.5cm]{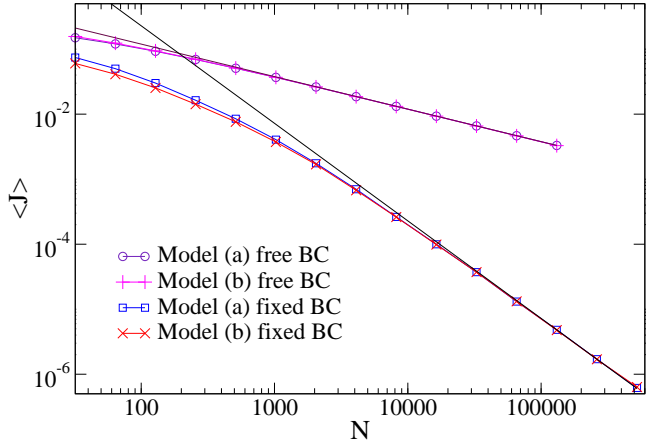}
\end{center}
\caption{(Color online). Plot of $\la J \ra$ versus $N$ 
for free BC ($n=0$) and fixed BC ($n=2$). 
Results are given for both models(a,b) of baths. 
The two straight lines correspond to the asymptotic expressions given
in Eqs.~(\ref{jfr},\ref{jfi}). We used parameters $m=1,~\Delta=0.5,~k=1,~
\g=1$, $T_L=2$, $T_R=1$ and $k_o=1$.  The error in the measurements is much smaller than
the size of the symbols.}
\label{fig1}
\end{figure}

\begin{figure}
\begin{center}
\includegraphics[width=8.5cm]{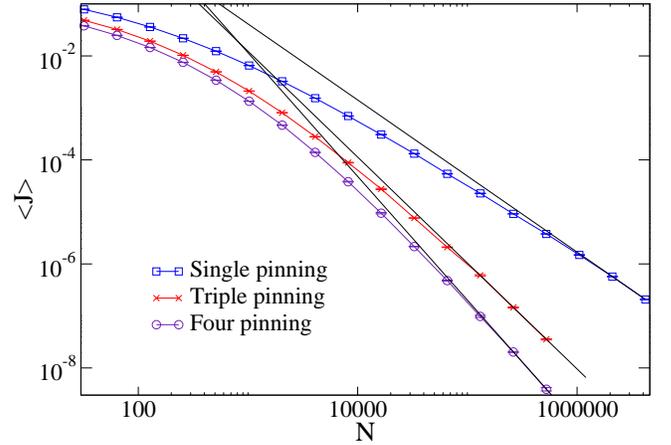}
\end{center}
\caption{(Color online). Plot of $\la J \ra$ versus $N$ for $n=1,3,4$ pinning
  centers for model(b). Parameters are same as in Fig.~(\ref{fig1})
  and for these parameters model(a) results are almost
  indistinguishable for $N>10^3$. The straight lines have slopes $-1.47$, $-2.03$ and $-2.38$.   
The error-bars shown are for disorder 
average and are of the same order as numerical errors.}
\label{fig2}
\end{figure}

For the choice of parameters $\gamma= (mk)^{1/2}$, the
imaginary part of $\Sigma'$ is the same for both bath models, and we
expect, for large system sizes, the actual values of the current to be
the same in both cases.  This can be seen in Fig.~(\ref{fig1})
where we show the system size dependence of the current for the
various cases. The current was evaluated numerically using
Eq.~(\ref{ocur1}) and averaging over many realizations $(\sim 4-100)$.
We also show the  exact asymptotic forms for
the current which we will discuss later.     
Note that for free BC, the exponent $\alpha=1/2$ settles to its
asymptotic value at  relatively small values ($N\sim10^3$)
while, with pinning,  we need to examine much longer chains ($N \sim
10^5$).
We also find that the presence of a single
pinning centre in the IDHC is sufficient to change the value of $\alpha$
from $1/2$  to $-1/2$ [see Fig.~(\ref{fig2})].
These results clearly show that, for both 
models(a,b), the  exponent $\alpha$ is the same and is controlled
by the presence or absence of pinning in the IDHC.

Next we try to better understand  the above results. As mentioned
before  only modes $\om \stackrel{<}{\sim} \om_d$ are
involved in conduction.  
It was noted in \cite{dhar01} that in this low frequency regime we can
approximate $\la \mT_N(\om) \ra$ by the
transmission coefficient of the ordered chain $\mT_N^O(\om)$. We then obtain
\bea
\la J \ra \sim (T_L-T_R) \int_{0}^{\om_d} \mT_N^{O}(\om)
d\om~.
\label{asyCur}
\eea 
For model(a), $\mT_N^O$ in
the limit $N \to \infty$  is effectively given by \cite{roy08}:    
\bea
\mT^O(\om)=\f{ \g \om^2
\sqrt{4mk-m^2\om^2}}{{k'}^2+(\g^2+m(k-k'))\om^2}~.
\eea
We then find, for free BC ($k'=0$)  $\mT^O(\om) \sim 1 $ while
for fixed BC ($k'\neq 0$), $\mT^O(\om) \sim \om^2$. Using
Eq.~(\ref{asyCur}) then immediately gives the asymptotic $N$
dependence for the two BCs'.
Our results are valid  even in the
weak coupling limit $\g << 1$ and this means that the result 
given by MI for model(a) in the weak coupling limit is
incorrect. Our numerics supports this conclusion. 
We also compute the transmission coefficient of the ordered chain (as $N \to \infty$) in the presence of a single pinning at one boundary,
\bea
 \mT^O_\infty(\om)&=&\f{2 \g \om^2 \sqrt{4mk-m^2\om^2}}{\sqrt{4\om^4\Lambda^2+k'\Omega(k'\Omega+4\om^2\Lambda)}},\\
{\rm with}~~~\Lambda&=&\g^2+km,~~~\Omega=  k'-m\om^2,\nn
\eea
 and we again find $\mT^O \sim
\om^2$, for $k'\neq 0$. This confirms our numerics that the asymptotic $N$ dependnce of $\la J \ra$ is analogous for the IDHC with single or double pinning centres.   
For  model(b), the transmission
coefficient of the ordered chain, pinned at 
the two boundary sites with $k_o=k'-k$ is given effectively by (as $N\to \infty$):
\bea
\mT^O(\om)=\f{2k^2\sin^2 q}{2k^2\sin^2 q+k_o^2}~,
\eea
where $\om=2(k/m)^{1/2} \sin(q/2)$. As expected, for $k_o=0$ we have $\mT^O=1$ 
while for $k_o\neq 0$, $\mT^O \sim \om^2$. 
The above qualitative analysis thus shows that 
the effect of  introducing pinning potentials is to pinch the band of
conducting modes (between $0-\om_d$) from
the zero frequency side and thus lower $ \la J \ra$.

Our asymptotic analysis also allows us to make predictions, on the
dependence of $\la J\ra$, on various system parameters such as mass variance, spring
constant etc. Here we denote ${\la J\ra}_{Fr}$ for $\la J\ra$ in the absence of pinning while ${\la J \ra}_{Fi}$ represents $\la J\ra$ in the presence of double pinnings at the boundaries. From Eq.~(\ref{asyCur}) and the forms of $\mT^O(\om)$ in
various cases we get:
\bea
{\la J \ra}_{Fr}&=& A ~c~\f{k_B (T_L-T_R)}{\pi }\Big(\f{k m }{N \sigma^2}\Big)^{1/2}  \label{jfr}\\
{\la J \ra}_{Fi} &=& A'~ c'~  \f{k_B (T_L-T_R) }{\pi }\Big(\f{k m
}{N \sigma^2 }\Big)^{3/2}~ \label{jfi},  
\eea
where $c=2\gamma(mk)^{1/2}/({\gamma}^2+mk),~1$ for model(a), model(b)
respectively. For fixed boundaries we have
$c'=\g (mk)^{1/2}/{k_o}^2,~ mk/{k_o}^2$ for model(a), model(b) respectively.  
$A, A'$ are constant numbers. 
We find that for model(b) our numerical results agrees with an exact
expression for 
$\la J\ra _{Fr}$ due to Papanicolau (apart from a factor of $2 \pi$)
and this gives  
$A=\pi^{3/2} \int_{0}^{\infty}dt~ {[t \sinh(\pi t)]}/
{[(t^2+1/4)^{1/2} \cosh^2(\pi t)]} \approx 1.08417$ (see
[\onlinecite{verheggen79}]). 
We note that this  differs from the expression  given in \cite{Matsuda70}.
For fixed boundaries we find numerically that $A' \approx 17.28$ and
the fit is shown in Fig.~(\ref{fig1}). 
Based on our analytical and numerical results, we believe that the
expression  in [\onlinecite{verheggen79}] is in error by a $2 \pi$ factor. 

Till now, using numerical results and heurestic
arguments,  we have arrived at the result that for a IDHC, in the
absence of any pinning potential $\alpha=1/2$ while the presence of 
one or two pinned sites changes the exponent to
$\alpha=-1/2$. This is true  both for white noise and Rubin's bath. It
is natural to now ask as to what happens in the presence of more
number of pinning centers. It is expected that more pinning centers
will lead to enhanced scattering of low frequency phonons and decrease
the heat current but it is not obvious  as to whether the exponent
$\alpha$ changes. 
For a finite fraction of sites on the lattice having 
pinning potentials, it is known that $\la J \ra \sim 
  e^{-cN}$ \cite{DharLeb08}. Here we investigate 
the case with a finite number, say $n$, of pinning sites. 
Numerically it becomes difficult to determine $\alpha$  for
$n>4$  as, with more pins, the  heat current becomes very small at large
system sizes and  numerical
errors become significant. In Fig.~(\ref{fig2}) we show numerical
results for $n=3,4$, where the extra pinning potentials with $k_o=1$
are placed in the bulk of the chain with equal separations. We find $\alpha \approx
-1.03, -1.38$ respectively for $n=3,4$, which
are clearly different from the $n=1,2$ value $\alpha=-0.5$. 
Let us now see what our earlier heuristic arguments give, for $n=3$.
We again find that the low frequency behaviour of $\Delta_N(\om)$
are similar for the disordered and ordered lattices. Let us therefore
find the form of $\Delta_N$ for the ordered case. Let $N=2M+1$ with
the $1^{\rm st}$, $(M+1)^{\rm th}$ and $N^{\rm th}$ sites being pinned. 
Except
for $\hat{T}_{M+1}=\hat{T}'$ all the other $\hat{T}_l$s' are identical and given by  $\hat{T}$,
say. If we denote $\hat{D}_N=\hat{T}^N$ and ${\hat{D}'}_N=\hat{T}^M
\hat{T}' \hat{T}^M= \hat{D}_M \hat{T}' \hat{D}_M $, then using the
fact that for the ordered lattice 
$D_{1N}=\sin q(N+1)/\sin(q)$,
where $\cos(q) = 1- m \om^2/(2k)$,  
and carrying out the matrix multiplications above we find that at low
frequencies $D'_{1N}$ is larger than $D_{1N}$ by a  factor  $\sim
1/\sin(q) \sim 1/\om$. This means
that $\mT_N$ for the $3$-pin case will have an extra factor of
$\om^2$ compared to the $2$-pin case. Correspondingly one
expects, using Eq.~(\ref{asyCur}), an exponent 
$\alpha=-3/2$. The argument can be extended to the case of $n \ge 2$ pins
(two of which are in the boundaries) in which case we get 
\bea
\alpha =3/2-n~.
\eea
Our numerical results for $n = 3,4$ (see Fig.~(\ref{fig2})) are consistent with this
prediction though we are not able to verify the precise value of the exponent. 

Finally we note that the calculation by CL \cite{Casher71} for the lower bound
on current, in the case of  two pinning centers (fixed boundaries) in model(a),
can be extended to the case with more pins.
The argument by CL consists in evaluating $\la D_{1N}^2 \ra
$ by looking at the disorder averaged direct product $\la \hat{D}\otimes
\hat{D} \ra =\prod_l \la \hat{Q}_l \ra $ where  $\hat{Q}_l =\hat{T}_l \otimes \hat{T}_l$. In the CL case
$\la \hat{Q}_l \ra =\hat{Q} $ for all $\hat{Q}$ and an analysis of the eigenvalues of $\hat{Q}$
led to the result $\la D_{1N}^2 \ra \sim e^{c N \om^2}$.     
In our case, say for the case of $n=3$ with an additional pinning at site
$l=M+1$,  $\la \hat{Q}_{M+1} \ra =\hat{Q}'$ is different and we have 
$\langle \hat{D} \otimes \hat{D} \rangle= \hat{Q}^M\hat{Q}'\hat{Q}^M$.   
A careful analysis of this then gives  
$ \la D_{1N}^2 \ra  \simeq \om^{-2}e^{cN \om^2}$.
Using this in  Eq.~(\ref{ocur1})  gives 
$ {\la J \ra}_3 \ge  C \int_0^{\infty} d \om \om^4 e^{-c N 
\om^2}  \sim  O[N ^{-{5}/{2}}]$,  
where $c$ and $C$ are constants. 
In general we get ${\la J \ra}_n \ge  O[N^{-n+\f{1}{2}}]$ for $2\le n
<< N$.

Quantum case: For a Hamiltonian of the form of Eq.(\ref{ham}) where
now $\{x_l,p_l\}$ are Heisenberg  operators, the steady state quantum
heat current through the IDHC in the linear response regime is given
by: 
\bea
J_q=\f{k_B(T_L-T_R)}{4\pi}\int_{-\infty}^\infty d \om
\mT_N(\om)\big(\f{\hbar \om}{2k_BT}\big)^2{\rm cosech}^2\big(\f{\hbar \om}{2k_BT}\big),\label{ocur2}\nn
\eea  
where $\mT_N(\om)$ is same as Eq.(\ref{ocur1}) and
$T=(T_L+T_R)/2$. Following our derivation for the classical system we
see that the asymptotic $N$ dependence of $\la J_q \ra$ is determined
by $\mT_N(\om)$ which is here exactly the same as the classical case. 
For any fixed temperature, however small, at suficiently large system
sizes we will have $\hbar \om_d << k_BT$, and hence within this
cut-off frequency the factor $(\hbar \om/k_B T)^2  {\rm cosech}(\hbar
\om/k_B T)^2 \to 1$. Hence for large system sizes we always get the
classical result. The approach to the asymptotic behaviour though will
be different.

Discussion: In real experiments heat baths  usually have a finite
bandwidth making the noise correlated, as in Rubin's model. Here we
have shown that for heat conduction in the IDHC these noise correlations do
not affect the  exponent $\alpha$ (note that a bath for which
$\Sigma(\om)$ depends  nonlinearly on  $\om$ at small frequencies can affect $\alpha$). 
We have elucidated the role of boundary conditions and shown that the
actual value of $\alpha$ depends on the number of pinned sites. 
Our results are also valid for bond disorder. We have provided
explicit expressions for the currents which, apart from giving the system
size-dependence, also give the dependence on various other parameters
such as mass variance, coupling to baths etc. We also emphasize that
heat conduction through IDHC is non-diffusive.
Our  physical understanding is as follows.
In the presence of mass or bond disorder phonons are scattered
coherently giving rise to localization and low transmission.
Long wavelength phonons with $\om \stackrel{<}{\sim} \om_d$
[see Eq.(1)] are relatively unaffected and 
dominate  heat conduction in such  disordered 
materials. Now the introduction of pinning  centers  causes strong 
scattering of even the low frequency modes and, as we
have shown, significantly reduces the current. We obtain the
surprising and nontrivial result that  the exponent $\alpha$ giving the
system size dependence of current changes linearly with the number
of pinning centers. 
There are now experimental measurements of heat conduction in
one-dimensional systems such as nanotubes and 
nanowires \cite{chang06,angel98} and molecular wires \cite{wang07}.
At low temperatures  one can neglect anharmonic
effects and it will be interesting to see if our prediction of the
strong reduction of heat current, by substrate potentials at localized
points on a disordered wire, can be observed.
While our results are for a simple classical model we expect 
the effect of pinning to be quite generic and should be true for
systems with more complicated phonon dispersions.
It will be interesting to see the  effect of pinning potentials in
heat conduction in two and three dimensions.

AD thanks David Huse for useful discussions.

\end{document}